\documentstyle[twoside]{article}
\input{epsfig.sty}
\catcode`\@=11
\long\def\@makefntext#1{
\protect\noindent \hbox to 3.2pt {\hskip-.9pt  
$^{{\eightrm\@thefnmark}}$\hfil}#1\hfill}		%CAN BE USED 

\def\thefootnote{\fnsymbol{footnote}}
\def\@makefnmark{\hbox to 0pt{$^{\@thefnmark}$\hss}}	%ORIGINAL 
	
\def\ps@myheadings{\let\@mkboth\@gobbletwo
\def\@oddhead{\hbox{}
\rightmark\hfil\eightrm\thepage}   
\def\@oddfoot{}\def\@evenhead{\eightrm\thepage\hfil
\leftmark\hbox{}}\def\@evenfoot{}
\def\sectionmark##1{}\def\subsectionmark##1{}}

\oddsidemargin=\evensidemargin
\renewcommand{\thefootnote}{\fnsymbol{footnote}}

\newcounter{sectionc}\newcounter{subsectionc}\newcounter{subsubsectionc}
\renewcommand{\section}[1] {\vspace{12pt}\addtocounter{sectionc}{1} 
\setcounter{subsectionc}{0}\setcounter{subsubsectionc}{0}\noindent 
	{\tenbf\thesectionc. #1}\par\vspace{5pt}}
\renewcommand{\subsection}[1] {\vspace{12pt}\addtocounter{subsectionc}{1} 
	\setcounter{subsubsectionc}{0}\noindent 
	{\bf\thesectionc.\thesubsectionc. {\kern1pt \bfit #1}}\par\vspace{5pt}}
\renewcommand{\subsubsection}[1] {\vspace{12pt}\addtocounter{subsubsectionc}{1}
	\noindent{\tenrm\thesectionc.\thesubsectionc.\thesubsubsectionc.
	{\kern1pt \tenit #1}}\par\vspace{5pt}}
\newcommand{\nonumsection}[1] {\vspace{12pt}\noindent{\tenbf #1}
	\par\vspace{5pt}}

\newcounter{appendixc}
\newcounter{subappendixc}[appendixc]
\newcounter{subsubappendixc}[subappendixc]
\renewcommand{\thesubappendixc}{\Alph{appendixc}.\arabic{subappendixc}}
\renewcommand{\thesubsubappendixc}
	{\Alph{appendixc}.\arabic{subappendixc}.\arabic{subsubappendixc}}

\renewcommand{\appendix}[1] {\vspace{12pt}
        \refstepcounter{appendixc}
        \setcounter{figure}{0}
        \setcounter{table}{0}
        \setcounter{lemma}{0}
        \setcounter{theorem}{0}
        \setcounter{corollary}{0}
        \setcounter{definition}{0}
        \setcounter{equation}{0}
        \renewcommand{\thefigure}{\Alph{appendixc}.\arabic{figure}}
        \renewcommand{\thetable}{\Alph{appendixc}.\arabic{table}}
        \renewcommand{\theappendixc}{\Alph{appendixc}}
        \renewcommand{\thelemma}{\Alph{appendixc}.\arabic{lemma}}
        \renewcommand{\thetheorem}{\Alph{appendixc}.\arabic{theorem}}
        \renewcommand{\thedefinition}{\Alph{appendixc}.\arabic{definition}}
        \renewcommand{\thecorollary}{\Alph{appendixc}.\arabic{corollary}}
        \renewcommand{\theequation}{\Alph{appendixc}.\arabic{equation}}
        \noindent{\tenbf Appendix \theappendixc #1}\par\vspace{5pt}}
\newcommand{\subappendix}[1] {\vspace{12pt}
        \refstepcounter{subappendixc}
        \noindent{\bf Appendix \thesubappendixc. {\kern1pt \bfit #1}}
	\par\vspace{5pt}}
\newcommand{\subsubappendix}[1] {\vspace{12pt}
        \refstepcounter{subsubappendixc}
        \noindent{\rm Appendix \thesubsubappendixc. {\kern1pt \tenit #1}}
	\par\vspace{5pt}}

\newcommand{\textlineskip}{\baselineskip=13pt}
\newcommand{\smalllineskip}{\baselineskip=10pt}

\def\eightcirc{
\begin{picture}(0,0)
\put(4.4,1.8){\circle{6.5}}
\end{picture}}
\def\eightcopyright{\eightcirc\kern2.7pt\hbox{\eightrm c}} 

\newcommand{\copyrightheading}[1]
	{\vspace*{-2.5cm}\smalllineskip{\flushleft
	{\footnotesize International Journal of Modern Physics A, #1}\\
	{\footnotesize $\eightcopyright$\, World Scientific Publishing
	 Company}\\
	 }}

\def\abstracts#1#2#3{{
	\centering{\begin{minipage}{4.5in}\baselineskip=10pt\footnotesize
	\parindent=0pt #1\par 
	\parindent=15pt #2\par
	\parindent=15pt #3
	\end{minipage}}\par}} 

\renewenvironment{thebibliography}[1]
	{\frenchspacing
	 \ninerm\baselineskip=11pt
	 \begin{list}{\arabic{enumi}.}
	{\usecounter{enumi}\setlength{\parsep}{0pt}
	 \setlength{\leftmargin 12.7pt}{\rightmargin 0pt} %FOR 1--9 ITEMS
	 \setlength{\itemsep}{0pt} \settowidth
	{\labelwidth}{#1.}\sloppy}}{\end{list}}

\newcounter{itemlistc}
\newcounter{romanlistc}
\newcounter{alphlistc}
\newcounter{arabiclistc}

\newcommand{\fcaption}[1]{
        \refstepcounter{figure}
        \setbox\@tempboxa = \hbox{\footnotesize Fig.~\thefigure. #1}
        \ifdim \wd\@tempboxa > 5in
           {\begin{center}
        \parbox{5in}{\footnotesize\smalllineskip Fig.~\thefigure. #1}
            \end{center}}
        \else
             {\begin{center}
             {\footnotesize Fig.~\thefigure. #1}
              \end{center}}
        \fi}

\newcommand{\tcaption}[1]{
        \refstepcounter{table}
        \setbox\@tempboxa = \hbox{\footnotesize Table~\thetable. #1}
        \ifdim \wd\@tempboxa > 5in
           {\begin{center}
        \parbox{5in}{\footnotesize\smalllineskip Table~\thetable. #1}
            \end{center}}
        \else
             {\begin{center}
             {\footnotesize Table~\thetable. #1}
              \end{center}}
        \fi}

\def\@citex[#1]#2{\if@filesw\immediate\write\@auxout
	{\string\citation{#2}}\fi
\def\@citea{}\@cite{\@for\@citeb:=#2\do
	{\@citea\def\@citea{,}\@ifundefined
	{b@\@citeb}{{\bf ?}\@warning
	{Citation `\@citeb' on page \thepage \space undefined}}
	{\csname b@\@citeb\endcsname}}}{#1}}

\newif\if@cghi
\def\cite{\@cghitrue\@ifnextchar [{\@tempswatrue
	\@citex}{\@tempswafalse\@citex[]}}
\def\citelow{\@cghifalse\@ifnextchar [{\@tempswatrue
	\@citex}{\@tempswafalse\@citex[]}}
\def\@cite#1#2{{$\null^{#1}$\if@tempswa\typeout
	{IJCGA warning: optional citation argument 
	ignored: `#2'} \fi}}

\def\pmb#1{\setbox0=\hbox{#1}
	\kern-.025em\copy0\kern-\wd0
	\kern.05em\copy0\kern-\wd0
	\kern-.025em\raise.0433em\box0}

\def\fnt#1#2{\footnotetext{\kern-.3em
	{$^{\mbox{\scriptsize #1}}$}{#2}}}

\def\fpage#1{\begingroup
\voffset=.3in
\thispagestyle{empty}\begin{table}[b]\centerline{\footnotesize #1}
	\end{table}\endgroup}

\def\runninghead#1#2{\pagestyle{myheadings}
\markboth{{\protect\footnotesize\it{\quad #1}}\hfill}
{\hfill{\protect\footnotesize\it{#2\quad}}}}
\font\tenrm=cmr10
\font\tenit=cmti10 
\font\tenbf=cmbx10
\font\bfit=cmbxti10 at 10pt
\font\ninerm=cmr9

\font\eightrm=cmr8

\def\qed{\hbox{${\vcenter{\vbox{			%HOLLOW SQUARE
   \hrule height 0.4pt\hbox{\vrule width 0.4pt height 6pt
   \kern5pt\vrule width 0.4pt}\hrule height 0.4pt}}}$}}

\renewcommand{\thefootnote}{\fnsymbol{footnote}}	%USE SYMBOLIC FOOTNOTE

\begin{document}

\runninghead{How to get less helium $\ldots$}
{How to get less helium $\ldots$}

\normalsize\textlineskip
\thispagestyle{empty}
\setcounter{page}{1}

\copyrightheading{}			%{Vol. 0, No. 0 (1993) 000--000}

\vspace*{0.88truein}

\fpage{1}
\centerline{\bf HOW TO GET LESS HELIUM} 
\vspace*{0.035truein}
\centerline{\bf AND MORE NEUTRINOS FROM BBN}
\vspace*{0.37truein}
\centerline{\footnotesize Xuelei Chen\footnote{Electronic Address: 
xuelei@pacific.mps.ohio-state.edu}}
\vspace*{0.015truein}
\centerline{\footnotesize\it Physics Department, the Ohio State University,
}
\baselineskip=10pt
\centerline{\footnotesize\it 174 W18th Ave., Columbus, OH~43210, 
USA}
\vspace*{10pt}

%\publisher{}{}

\vspace*{0.21truein}
\abstracts{
We discuss BBN in the presence of a non-minimally coupled quintessence
model. In some of these models, the gravitational constant 
and cosmic expansion rate are smaller than standard model
predicts. The Helium abundance is then smaller, possibly 
resolve the marginal disagreement between theory and 
observation. Furthermore, the constraint on neutrino 
species may also be relaxed.}{}{}

\textlineskip			%) USE THIS MEASUREMENT WHEN THERE IS
\vspace*{12pt}			%) NO SECTION HEADING

\vspace*{1pt}\textlineskip	%) USE THIS MEASUREMENT WHEN THERE IS

\noindent

\textheight=7.8truein
\setcounter{footnote}{0}
\renewcommand{\thefootnote}{\alph{footnote}}

The predicted primordial $~^4\mathrm{He}$ abundance $Y$ increases with the 
expansion rate of the Universe during the big bang nucleosynthesis(BBN).
This has been used to put limits on the number of neutrinos,  
quintessence models, and other new physics. 

At present, there are still large systematic errors in measurement of 
the primordial helium abundance. Oliver and
Steigman\cite{Oliver-Steigman} 
obtained
$Y = 0.234 \pm 0.003 ({\rm stat.})$,
while Izotov and Thuan\cite{Izotov-Thuan} obtained a higher value 
$Y = 0.244 \pm 0.002 ({\rm stat.}).$
Obviously these two data sets are statistically inconsistent with each.
In this paper, I shall adopt adopt a midway value of
$Y_p = 0.239 \pm 0.005$, or, $0.229< Y_p < 0.249$ at 95\% C.L.

The helium abundance also depends on the baryon to photon ratio
$\eta$. We can determine $\eta$ by either measuring deuterium
abundance, or fit CMB data with varying cosmological parameters.
Burles and Tytler\cite{Burles-Tytler} found
D/H=$(3.3 \pm 0.25) \times 10^{-5}$, the 
lower bound on $\eta$ at $2\sigma$ level is 
$\eta_{10} \equiv 10^{10} \eta <6.3$. On the other hand, 
a recent analysis\cite{Jaffe} of the CMB data yields 
$\Omega_b h^2 =0.030 \pm 0.004$, or $\eta_{10} = 8.2\pm 1.0$.
If we assume that there are three neutrino
species, and adopt $\eta \approx 4.5$ as inferred from the deuterium
abundance, then the standard BBN 
$~^{4}\mathrm{He}$ abundance is in disagreement with the result
of Oliver and Steigman. It is in marginal agreement with 
our ``midway'' result, but still at the higher end. 
Even this ``midway'' limit is exceeded if  
the $\eta$ inferred from CMB is adopted(see Fig. 1).

Furthermore, in addition to the three standard model neutrinos, 
a sterile neutrino may be needed to explain the results from
neutrino oscillation experiments\cite{neutrino}. If either or both of 
these were confirmed, or if there are any other light particles in the 
Universe, the breach between theory and observation on
$~^{4}\mathrm{He}$ would become even wider.

Here, I show that with non-minimally coupled quintessence 
model\cite{ExtendedQuintessence,NMC}, the predicted helium abundance
could be lowered, thus alleviate the breach between theory and
observation. Alternatively, in such a model the cosmological bound
on neutrino number is relaxed, making room for a possible fourth neutrino. 

The action for the NMC model is given by
\begin{equation}
S=\int d^{4} x \sqrt{-g} \left[\frac{1}{2}F(Q) R-\frac{1}{2}  Q^{;\mu} 
Q_{;\mu} -V(Q) + L_{\rm fluid}\right],
\label{NMCeq}
\end{equation}
with $F(Q)=1-\xi(Q^2 - Q_0^{2})$, where $Q_0$ is a constant, and  
$V(Q)=V_{0} Q^{-\alpha}$.
It is known that in this model the cosmic expansion accelerates at
late time\cite{NMC}, 
consistent with the recent type Ia supernova measurement\cite{SNIa}.

In this model, the expansion rate of the Universe is
\begin{equation}
H^2 = \frac{1}{3F} 
\left(\rho_f + \frac{1}{2} \dot{Q}^2 +V(Q) - 3 H \dot{F}
\right),
\label{Heq}
\end{equation}
where $\rho_f = \rho_m +\rho_r$ is the density contribution from 
matter and radiation (including neutrino).
The change in expansion rate could be 
parametrized by a speed-up factor $\zeta$:
\begin{equation}
\zeta \equiv \frac{H}{\bar{H}} \approx \xi(Q_i^2 - Q_0^2).
\end{equation}

\begin{figure}[htbp]
\begin{center}
\epsfig{file=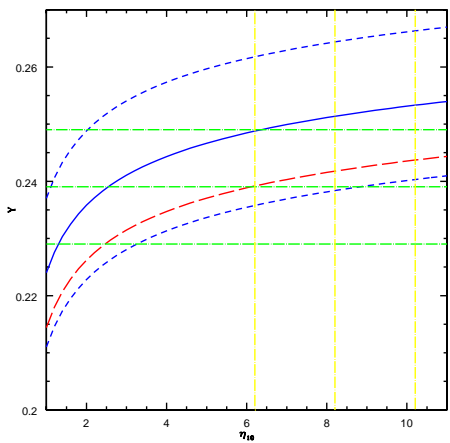,width=2.5in}
\caption{The Helium abundance as a function of $\eta_{10}$. The Solid
curve shows standard BBN result with three neutrinos, the two short-dashed
curves denotes the cases of two and four neutrinos respectively, and 
the long-dashed curve is the result of the NMC model discussed in text.
The three horizontal lines shows the
center value and 2$\sigma$ bound on observed helium abundance (``mixed
result''), while the three vertical lines indicate the center value and
2$\sigma$ bounds on $\eta$ from CMB.}
\end{center}
\label{Y-eta}
\end{figure}

The effect of a constant speed up factor on helium abundance has been 
investigated in the context of neutrino number limit. We have 
\begin{eqnarray}
Y&=&(0.2378+0.0073\ln\eta_{10})(1-0.058/\eta_{10})\nonumber\\
&&+0.013\Delta N_{\nu}+2\times 10^{-4} (\tau -887),
\end{eqnarray}
the speed up factor is related to neutrino number by
\begin{equation}
\zeta^{2} = 1+\frac{7}{43} \Delta N_{\nu}, 
\end{equation}
so we have 
\begin{equation}
\Delta Y = 0.08 (\zeta^2 -1) \approx 0.16 (\zeta -1 ).
\end{equation}

Thus, for example, in a model with $\alpha=10$, $\xi=0.004$, we find
$\zeta-1 \approx -0.06$, $\Delta Y \approx -0.96\%$. In terms of
neutrino number, this corresponds to a reduction of 
$\Delta N_{\nu} \approx 0.74$.  

The coupling constant $\xi$ is limited by solar system
experiment. 
For a model with $\alpha=10$, $\xi=0.004$, which is within solar
system limit, $\xi<0.022Q_{0}^{-1}$, 
the helium abundance is reduced by $0.96\%$. If there is indeed
a breach between the observed helium abundance and the standard BBN
theory as indicated by Olive and Steigman\cite{Oliver-Steigman}, 
it could be explained by this NMC model. Alternatively, such a shift
in $Y$ corresponds to a shift in neutrino number $\Delta N_{\nu}= -0.74$.
The current cosmological limit\cite{PDG} on neutrino number is $1.7 < N_{\nu} <
4.3$ at 95\% C.L., for the model described above, the upper
limit would be relaxed to 5.0, making sufficient room for a
non-standard model neutrino.  

\nonumsection{Acknowledgements}
\noindent
This work is supported by the US Department of Energy grant
DE-AC02-76ER-01545 at the Ohio State University.

\newcommand\AJ[3]{~Astron. J.{\bf ~#1}, #2~(#3)}
\newcommand\APJ[3]{~Astrophys. J.{\bf ~#1}, #2~ (#3)}
\newcommand\apjl[3]{~Astrophys. J. Lett. {\bf ~#1}, L#2~(#3)}
\newcommand\ass[3]{~Astrophys. Space Sci.{\bf ~#1}, #2~(#3)}
\newcommand\cqg[3]{~Class. Quant. Grav.{\bf ~#1}, #2~(#3)}
\newcommand\mnras[3]{~Mon. Not. R. Astron. Soc.{\bf ~#1}, #2~(#3)}
\newcommand\mpla[3]{~Mod. Phys. Lett. A{\bf ~#1}, #2~(#3)}
\newcommand\npb[3]{~Nucl. Phys. B{\bf ~#1}, #2~(#3)}
\newcommand\plb[3]{~Phys. Lett. B{\bf ~#1}, #2~(#3)}
\newcommand\pr[3]{~Phys. Rev.{\bf ~#1}, #2~(#3)}
\newcommand\PRL[3]{~Phys. Rev. Lett.{\bf ~#1}, #2~(#3)}
\newcommand\PRD[3]{~Phys. Rev. D{\bf ~#1}, #2~(#3)}
\newcommand\prog[3]{~Prog. Theor. Phys. {\bf ~#1}, #2~(#3)}
\newcommand\MeV{\mathrm{MeV}}

\nonumsection{References}

\end{document}  
\end